# Toroidal hollow-core microcavities produced by self-rolling of strained polymer bilayer films.

V. Luchnikov, K. Kumar, M. Stamm

*Leibniz Institute for Polymer Research Dresden, Hohe Str. 6,*

*D-01069 Dresden, Germany*

*Email:* luchnikov@ipfdd.de *or* luchnikov@yahoo.com

*Telefon:* +49(0351)4658272; Fax: +49(0351)4658281

*Short title:* Toroidal microcavities produced by self-rolling of strained polymer bilayer films.

Hollow-core toroidal micro-cavities are obtained by self-rolling of double-layer (polyvinyl pyridine/polystyrole) polymer films. Rolling of the bilayer is due to preferential swelling of the polyvinyl pyridine in water solution of dodecyl benzene sulfonic acid. The tube formation proceeds from a circular opening in the film made by photolithography or by mechanical scratching. The toroid equilibrium dimensions are determined by the balance of the elastic energy relaxation via the film scrolling and the work of the in-plane stretching that is due to increasing radius of the toroid. The principal features of the micro-toroid formation process are captured by a simple analytical model. The inner walls of the cavities can be made metal coated. To this aim, the polymer bilayer can be metallized by vacuum sputtering prior to lithographic patterning and rolling of the bilayer. The toroids with metallic inner surface are promising for the future research as IR-frequency range resonators.

**1.Introduction.**

Engineered micro- and nanocavities attract much attention as a means to trap light and matter in mesoscale volumes for the studies of confinement effects, such as inhibition or enhancement of spontaneous radiation rate [1] and depression of the melting point [2,3]. The cavities of toroidal shape are particularly interesting as quasi-one dimensional compartment with periodic boundary conditions. Recently, dielectric micro-rings produced of fused silica have been explored as high-finesse optical resonators, which have broad realm of possible applications in telecommunication and bio-sensing technologies [4,5].

In our paper we introduce a novel approach to fabrication of toroidal micro-cavities. The approach is based on the recently developed technique of micro- and nanotube fabrication via self-rolling of strained bilayer films [6-8]. The tubes prepared by this method are promising for many advanced applications including nano-syringes for intra-cellular surgery and nano-jet printing [9], X-ray and visible light waveguides [10]. The strain in the film, which creates the bending moment and rolling, can be caused by different factors such as the misfit of crystal lattices of the top and the bottom components of the bilayer [6,7], unequal thermal expansion of the film components [11], or different swelling of chemically distinct polymers in selective solvents [8]. The rolling proceeds from an opening in the film, created by photolithography route or by mechanical scratching. The opening provides the contact of an etching agent or a selective solvent with the lower layers of the system. Following the Ref.[9], we call these openings the lithographic windows (LW). The tubes explored in earlier works of our and other groups originated from straight LWs and had open ends. Here we show that closed looped tubes can be produced by self-rolling of a bilayer from circular openings. Contrary to the case of the straight tubes formation, which can roll until there is the film available, self-rolling of the toroidal tubes is self-constraining: it stops when the energy gain due to relaxation of the bending moment is compensated by the work on the in-plane stretching of the film caused by the increasing radius of the torus.

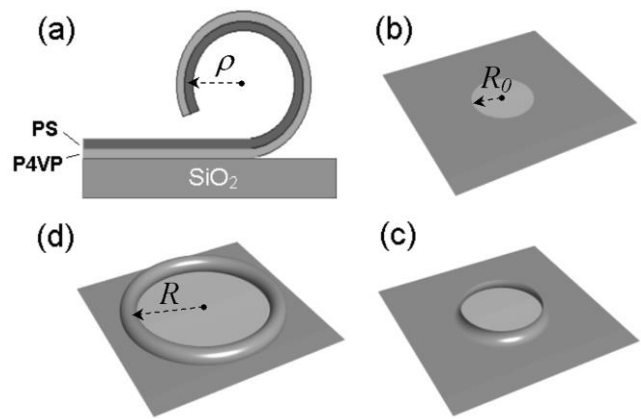

**Figure 1.** The scheme of formation of the rolled up toroidal micro tubes. (a) The structure of the polymer bilayer. (b) A circular lithographic window in the film produced by photolithography approach. (c),(d) initial and the final stages of the toroid formation.

## 2.Experiments.

The formation of the micro-toroids is schematically depicted on figure 1. In our fabrication scheme we explore the self-rolling effect produced by unequal swelling of the top and bottom components of bilayer polymer films [8] (figure 1a). The poly(4-vinyl pyridine) (P4VP) and the polystyrene (PS) layers were deposited from solutions in chloroform and toluene, respectively, by dip or spin-coating. The thickness of the layers (measured by ellipsometry) was $h_{p4vp}$=47 nm and $h_{PS}$=50nm. More details of the polymer bilayer film fabrication can be found in Ref.[8]. The film was patterned by irradiation of the sample through a photomask (a quartz slide with circular metallic features) with short-wave UV ($\lambda$ =254 nm, irradiation dose  W=4J/cm$^2$). UV-light of the short wavelength diapason leads to photo-crosslinking of the both polymers [12]. The pattern was developed by washing non-cured polymer in dichloromethane. In this way, the polymer films with circular openings (figure 1b) were obtained. Circular LWs were produced also by mechanical scratching. To this purpose the sample was mounted on a rotating stage, and a sharp needle was put by a micro-positioning system into contact with the sample surface. Patterned polymer bilayers were immersed in 2%- w.t. aqueous solution of dodecyl benzene sulfonic acid (DBSA) which forms supramolecular complexes with pyridine rings of P4VP [13] and increases the specific volume of the polymer. Since the PS layer is neutral in this solution, the bending moment arises which causes curling of the film and eventual tube formation (figure 1c,d).

SEM micrographs of micro-toroids produced by photolithography route are collected on figure 2. A defect-free 128 µm-wide toroid was formed by rolling of  3.3 µm-wide tube from a 100 µm-wide LW (figure 2a,b). The toroid has one completed shell, as can be deduced by comparing the width of the rolled layer  (14 µm)  and the circumference of the tube (approximately 10µm).  The surface of the tube is very smooth (see figure 2b).

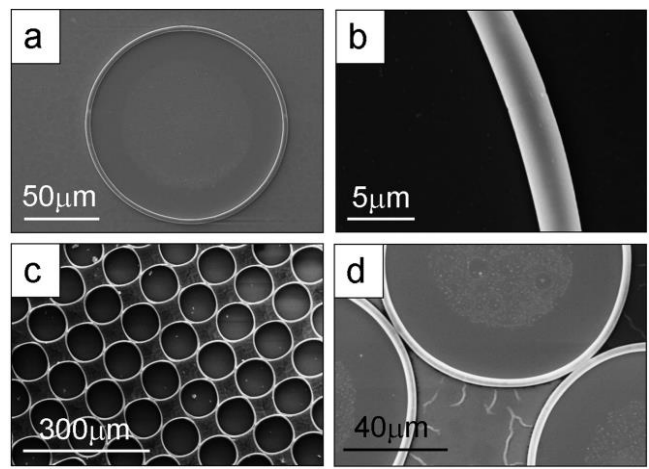

**Figure 2.** Toroidal tubes (SEM micrographs)    a), b): a single micro-toroid. c),d) an array of micro-toroids.

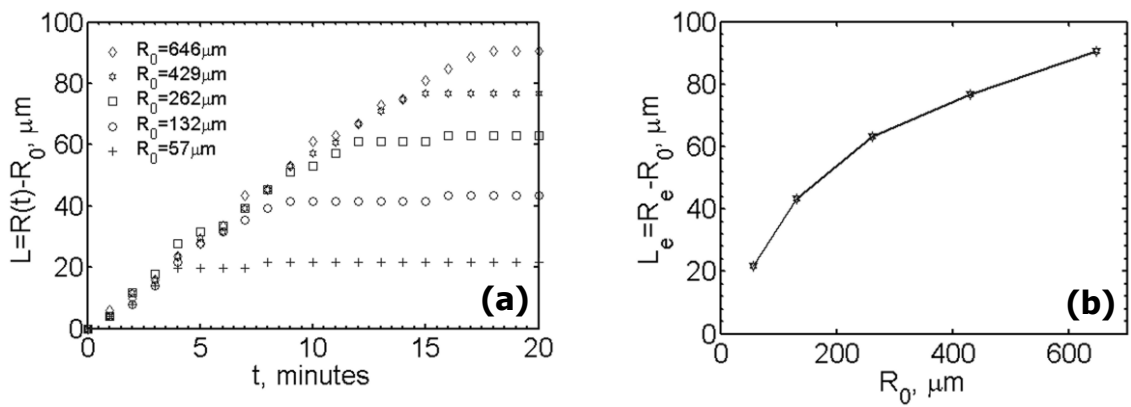

**Figure 3.** Characterization of the micro-toroid formation. a) The width of the rolled-up part of the bilayer as the function of time, for different radii of the lithographic window. b) The equilibrium width of the rolled-up part of the bilayer, as the function of the LW radius.

Photolithography permits to fabricate large arrays of toroids (figure 2c). If the distance between the neighbouring LWs is small, the toroids can touch each other, forming characteristic twin-tube junctions (figure 2d). In order to avoid the distortion of the toroid's shape by further rolling, the process can be stopped by removing the sample out of the solution and washing it in pure water. This operation can be done with high accuracy, because rolling proceeds sufficiently slow (typically, a few micrometers per minute).

As pointed above, the rolling of the toroidal tubes cannot proceed unlimitedly, as in the case of the straight tubes. The increasing radius of the toroid causes lateral strain in the film, such that the strain isolines are closed concentric circles centered at the middle of the LW. To investigate the kinetics of the toroid formation, we prepared by mechanical scratching a set of concentric LWs in the bilayer film and video-recorded the film rolling. Figure 3a shows the width of the rolled-up layer $L = R - R_0$ (see figure 1 for the notation) as the function of time for different LW radii. The rolling rate does not depend on the LW radius and amounts, for the given experimental conditions, $dL/dt \approx 5.3 \mu m \cdot \min^{-1}$. Apparently, it is the rate of swelling of the P4VP film that determines the tempo of toroid formation until the counter-forces originating in the lateral strain come into play. Rolling slows down when the width of the rolled part approaches a saturation value $L_e$ that depends on the radius of LW (see figure 3b). Larger radius of LW results in broader rolled-up stripe of the bilayer, since the effect of the curvature on the tube formation is smaller for wider LWs. This intuitive explanation of the experimental result is supported in the next section by a simple analytical model of the micro-toroid formation.

## 3. The model of the micro-toroid formation.

The process of micro-toroid formation can be understood on the qualitative level by considering the following simple model. Since the thickness of the film is much smaller than its radius of curvature, one can regard the film as a 2D-surface with the spontaneous curvature radius $\rho$. The contribution of the elastic energy to the system's free energy can be written as [14]:

$$F = \int \left[ \frac{1}{2} \kappa \left( \frac{1}{r_1} + \frac{1}{r_2} - \frac{2}{\rho} \right)^2 + \bar{\kappa} \left( \frac{1}{r_1 r_2} \right) \right] dA + \frac{1}{2} \int u_{\alpha\beta} \sigma_{\alpha\beta} dA \qquad (1)$$

Here $\kappa$ and $\bar{\kappa}$ are the mean and the Gaussian curvature elastic constants, respectively, $u_{\alpha\beta}$ and $\sigma_{\alpha\beta}$ are the two-dimensional deformation and stress tensors, calculated in appropriate curvilinear coordinates [16]. The integrals are taken over the whole surface of the film. The equilibrium form and dimensions of the tube released from the substrate can be found in principle by variation of the free energy functional (1). This a complicated problem is out of scope of our present paper. However, the model became easily tractable if we consider toroids originating from the LWs of large radii. In this case, it is reasonable to assume that $\rho << L$, i.e. the width of the rolled part of the film is much larger than the radius of the tube, because at the beginning of rolling the effect of the LW boundary curvature is small. Also, we suppose that the width of the rolled-up stripe of the bilayer is small compared to the radius of the LW, $L << R_0$. This condition is fulfilled if the stretching stiffness of the film is sufficiently large. Finally, for the thin looped tube one can always assume $r_2^{-1} << r_1^{-1} \approx \rho^{-1}$, i.e. the maximal curvature of any element of the tube's surface is much larger than the minimal curvature, that is approximately equal to the spontaneous curvature. Let $Y$ be the effective Young modulus of the double layer film [17]. Then, the difference of the free energy of the film in the rolled and the unrolled states can be easily integrated (see Appendix A) and reads, in the dimensionless units $\Delta \tilde{F} = \Delta F / (\pi Y R_0^2)$, $x = L / R_0 << 1$ as:

$$\Delta \tilde{F}(x) = -A \cdot \left[ x^2 + 2x \right] + (1+x)^2 \ln(1+x) - x - \frac{3}{2} x^2 \approx -A \cdot \left[ x^2 + 2x \right] + \frac{x^3}{3} \qquad (2)$$

where $A = 3\kappa / (2\rho^2 Y)$. The function $\tilde{F}(x)$ has the minimum at $x_e = A + \sqrt{A^2 + 2A}$ (see figure 4). Self-consistency of the condition $x << 1$ requires that $A << 1$, and, up to the leading term, the position of the minimum is $x_e \approx \sqrt{2A}$. Thus, the equilibrium number of shells of the toroidal tube is:

$$n_e = L_e / (2\pi\rho) = x_e R_0 / (2\pi\rho) = \frac{R_0}{2\pi\rho^2} \sqrt{\frac{3\kappa}{Y}} \qquad (3)$$

In this paper, we do not make direct comparison of our model with experiment, since the mechanical properties of the P4VP films forming supramolecular complexes with DBSA are unknown. Nevertheless, the model qualitatively explains the following facts about the toroidal tube formation: a) rolling stops after reaching a certain width of the rolled-up part of the bilayer (apparently, this equilibrium width corresponds to the minimum of the free energy), and b) the equilibrium width of the rolled-up film is larger for the circular LWs of larger radii, in accordance with the experiment (see figure 3b). However, the proportionality of $L_e$ and $R_0$ is not linear, as suggested by the formula (3). This is an indication that a more detailed theory need to be developed. Finally, the model supports the intuitive expectation that the number of the shells (or, equivalently, the width of the rolled-up bilayer) is larger for the films with higher values of the bending modulus (hence, providing larger energy gain via scrolling) and smaller stiffness of the in-plane stretching (that is, causing less constraint to rolling as the diameter of the toroid increases). This tendency is illustrated on the figure 4 by plotting several free energy graphs for the increasing values of the parameter *A*.

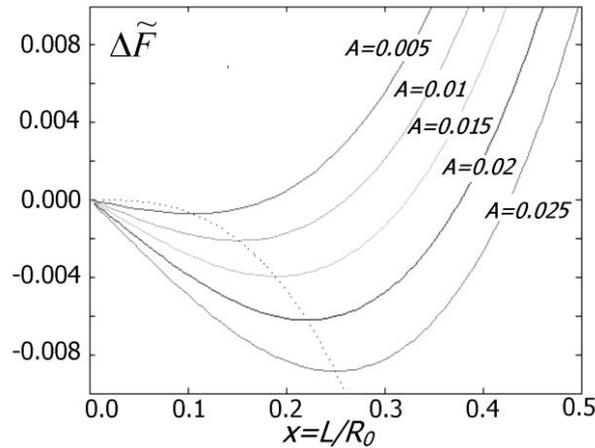

**Figure 4.** Free energy of the micro-toroid as the function of the width of the rolled-up layer, for different values of the parameter *A*. (in dimensionless units). The dotted line indicates the position of the function minimum.

### 4. Conclusion.

We have introduced fabrication of hollow-core toroidal micro-cavities by the self-rolling of strained polymer bilayer films. A few micrometer wide tubes are looped in the few dozens wide toroids. The formation of the toroids is self-constrained: their equilibrium width is determined by the balance of the free energy gain due to relaxation of the bending moment and the work of the film stretching. Our analytical model of the micro-toroid formation qualitatively correct predicts the existence of the equilibrium size of the micro-toroid. However, a more detailed theory should be elaborated. In particular, coupling of bending and stretching should be taken into account while calculating the free energy of the rolling bilayer. The necessity of such a coupling in the multi-layer films is obvious from the fact that stretching the film in one direction will produce bending of the film in the orthogonal plane, as soon as the Poisson ratios of the materials of the layers differ. Also, the nonlinear stress-strain relation for the polymer film should be taken into account. These issues will be addressed in the future paper.

Hollow-core toroidal micro-cavities produced by self-rolling approach can be interesting for a number of advanced applications. In a few preliminary experiments, we modified the hidden walls of the tubes by magnetron sputtering of gold on the surface of the polymer film, prior to film rolling. The circular LWs were made by mechanical scratching. Upon rolling, the micro-toroids with metallized inner surface were obtained. On the other hand, the hollow-core micro-capillaries with metallized inner surface are known to function as waveguides for IR-radiation [18]. Thus, the toroidal rolled-up cavities with the metallized inner walls may be interesting for the development of micro-resonators for the IR frequency diapason. Another possible field of application of the self-rolled micro-toroids is encapsulation of micro-particles of different nature (quantum dots, living cells *etc*). During rolling, the particles can be the captured from the surrounding solution and enclosed in the toroidal geometry. This may provide new opportunities for the studies of self-assembly phenomena in confined space [19].

**Appendix A**

General expression for the free energy functional of the film (1) can be simplified in frames of the following assumptions: a) $\rho << L$, b) $L << R_0$, c) $r_2^{-1} << r_1^{-1} \approx \rho^{-1}$ (see the definitions of the symbols in the main text). Below we use the cylindrical system of coordinates.

Energy relaxation via rolling. Before rolling, the curvature radii of the film aspire to infinity, and the first term in the free energy functional (1) can be written as:

$$F_i^{bend} \approx \int \left[ \frac{1}{2}\kappa \left( \frac{1}{\infty} + \frac{1}{\infty} - \frac{2}{\rho} \right)^2 + \bar{\kappa} \left( \frac{1}{\infty} \right) \right] dA = \frac{4}{2}\kappa \rho^{-2} \int dA \qquad (A1)$$

After scrolling, the free energy term associated with bending became, according to the assumption (c):

$$F_f^{bend} \approx \int \left[ \frac{1}{2}\kappa \left( \frac{1}{\infty} + \frac{1}{\rho} - \frac{2}{\rho} \right)^2 + \overline{\kappa} \left( \frac{1}{\infty \rho} \right) \right] dA = \frac{1}{2}\kappa \rho^{-2} \int dA \qquad (A2)$$

Thus, the free energy difference due to bending moment relaxation is:

$$\Delta F_{bend} = F_f^{bend} - F_i^{bend} = -\frac{3}{2}\kappa \rho^{-2} \int_{R_0}^{R} 2\pi r \, dr = -\frac{3\pi \kappa}{2\rho^2}\left( R^2 - R_0^2 \right) \qquad (A3)$$

<u>Work on the film stretching due to rolling.</u> Let be $r$ the radius of a circular element of surface of the width $dr$, prior to rolling (see figure A1). By the assumption (b), any element of surface has, *after rolling,* approximately the same distance to the centre: $r \to R$. There, the elongation of the surface element upon rolling can be written for all elements of surface as $\Delta l = 2\pi(R - r)$. The work of the in-plane stretching of is $dF_s = (1/2)Y \frac{\Delta l^2}{l} dr = \pi Y \frac{(R-r)^2}{r} dr$. Here, $Y$ is the effective stretching stiffness of the bilayer film. The elastic energy stored in the film upon rolling is:

$$\Delta F_{stretch} \approx \pi Y \int_{R_0}^{R} \frac{(R-r)^2}{r} dr = \pi Y \left[ R^2 \ln \frac{R}{R_0} - 2R(R - R_0) + \frac{1}{2}\left( R^2 - R_0^2 \right) \right] \qquad (A4)$$

Combining (A3) and (A4) gives:

$$\Delta F = \Delta F_{bend} + \Delta F_{stretch} = -\frac{3\pi \kappa}{2\rho^2}\left( R^2 - R_0^2 \right) + \pi Y \left[ R^2 \ln \frac{R}{R_0} - 2R(R - R_0) + \frac{1}{2}\left( R^2 - R_0^2 \right) \right] =$$

$$= \pi Y R_0^2 \left\{ -\frac{3\kappa}{2\rho^2 Y}\left[ \left( \frac{R}{R_0} \right)^2 - 1 \right] + \left( \frac{R}{R_0} \right)^2 \ln \frac{R}{R_0} - 2\frac{R}{R_0}\left( \frac{R}{R_0} - 1 \right) + \frac{1}{2}\left[ \left( \frac{R}{R_0} \right)^2 - 1 \right] \right\} \qquad (A5)$$

Let us introduce the dimensionless units, $\qquad \Delta \widetilde{F} = \dfrac{\Delta F}{\pi Y R_0^2} \quad$ and $\quad x = \dfrac{R - R_0}{R_0} = L/R_0$

Note that $\quad \dfrac{R}{R_0} = 1 + x$

Then, (A5) can be rewritten as:

$$\Delta \widetilde{F} = -A\left( 2x + x^2 \right) + (1 + x)^2 \ln(1 + x) - 2x(1 + x) + \frac{1}{2}\left( 2x + x^2 \right) \qquad (A6)$$

where $\qquad A = \dfrac{3\kappa}{2Y\rho^2}$

Since $x << 1$ by the assumption (b),   $\ln(1+x) \approx x - \dfrac{x^2}{2} + \dfrac{x^3}{3} + ...$

To the 3 eldest terms by $x$, the free energy difference of the rolled and unrolled states of the strained bilayer reads:

$$\Delta \widetilde{F} = -A\left(2x + x^2\right) + \frac{x^3}{3} \tag{A7}$$

The function has the minimum at $x_e = A + \sqrt{A^2 + 2A} \approx \sqrt{2A}$  (marked by dashed line on the figure 4).  Thus, the equilibrium width of the rolled-up layer is:

$$L_e \approx \frac{R_0}{\rho} \sqrt{\frac{3\kappa}{Y}} \tag{A6}$$

Since $x << 1$, there should be $A << 1$, i.e $\kappa << Y\rho^2$. This means that the model is self-consistent if bending stiffness of the film is sufficiently small.